\documentclass[a4paper,twocolumn,pra]{revtex4}
\usepackage{amsmath}
\usepackage{graphicx}
\usepackage{epstopdf}

\begin{document}

\title{Telegraph noise effects on two charge-qubits in double quantum dots}

\author{A. Ayachi$^{1}$, W. Ben Chouikha$^{1}$, S. Jaziri$^{1,2}$  and R. Bennaceur$^{1}$}

\affiliation{$^{1}$
Laboratoire de Physique de la Mati\`{e}re Condens\'{e}e. Facult\'{e} des Sciences de Tunis, Universit\'{e} de Tunis el Manar, 2092 campus universitaire de tunis, Tunisia
\\
$^{2}$ Facult\'{e} des Sciences de Bizerte,Universit\'{e} de Carthage  7021 Zarzouna, Bizerte, Tunisia}

\begin{abstract}
We analyze theoretically the decoherence of two interacting electrons in a double self-assembled quantum dot due to a random telegraph noise. For this purpose we have examined the pure dephasing rate by evaluating the decoherence factor. This latter has been shown to be different from that calculated within the Gaussian approximation in the strong coupling regime. In order to determine the influence of the random telegraph noise on the entanglement of the system states, the concurrence, the populations and the entropy are evaluated as well. Our results show that telegraph noise can severely impact the coherence of charges qubits.

\end{abstract}

\maketitle

\vspace{1cm}

\qquad In connection with quantum information processing, semiconductor quantum dots seem very promising to implement qubits and to achieve a quantum computer. Sundry experiments have already highlighted the quantum properties of these devices \cite{{key1}, {key2}, {key3}}. Several designs for the physical realization of qubit have been suggested and a wide variety of experiments have demonstrated the possibility of controlling the spin \cite{{key2}, {key4}} and the charge states \cite{{key3}, {key5}} of the confined electrons in the quantum dots. It is widely accepted that spin qubits hold a great promise in the long term on account of the large spin decoherence time characteristic, nevertheless, charge-based qubits are receiving increasing interest as well. Indeed, employing the charge degree of freedom of electrons rather than their spin brings substantial practical advantages since the experimental techniques for measuring and manipulating electron charge are extremely developed and there is no need for local control of magnetic fields and all the operations are accessible involving just low electric fields \cite{{key3}, {key5}}.

An attractive platform to study the quantum control of a charge qubit is the system of lateral quantum dots \cite{{key6}, {key7}}. A drawback to their use in quantum information is that they are coupled to the external degrees of freedom which leads to decoherence.
Thereby, charge qubits in a double quantum dot undergoes various decoherence mechanisms caused by the charge motion. Given that the qubit states are defined through the position of a mobile electron, i.e. the logical states correspond to the electron being on the right or left dot, the amendment of this state implies an electron jump from one quantum dot to another. Such motion can couple to external degrees of freedom such as phonons, impurity and electromagnetic fluctuations. Recent theoretical and experimental studies have been developed to investigate the effect of phonons in semiconductor quantum dot as a source of dephasing accompanying dissipation \cite{{key7}, {key8}, {key9}}. The effect of electromagnetic fluctuations has been widely examined \cite{{key9}, {key10}}.

Recently, evidence for noise due to the low frequency fluctuations has been observed in both Josephson junction structures \cite{key11} and in semiconductor self-assembled quantum dots \cite{{key12}, {key13}, {key14}}.
Several mechanisms have been proposed to explain these fluctuations in terms of: a fluctuating background charges, or a structural dynamic defects, or a charge traps \cite{{key15}, {key16}, {key17}, {key18}, {key19}, {key20}, {key21}}, and can hence be modeled as two-level systems; fluctuators. A random switching of one fluctuator between their two level tunneling states produces a random telegraph noise $(RTN)$. If many of these fluctuators are appropriately superimposed, they can lead to $1/f$ noise. Systems showing $1/f$ noise have been extensively studied in several papers \cite{{key19}, {key20}, {key21}, {key22}, {key23}, {key24}, {key25}, {key26}}.
Unfortunately, on account of the miniaturization of quantum dots and the development of techniques for their manufacture, the recent experiments prove the presence of only one or a few number of fluctuators \cite{key27}. Therefore, numerous studies have been devoted to the investigation of the decoherence in Josephson qubits due to RTN as a source of dephasing through evaluating the phase memory decay using first the spin-boson model \cite{{key28},{key29}, {key30}} and then the spin-fluctuator model \cite{{key16}, {key22}, {key31}}. Recent works point out that the last model is more appropriate than the Gaussian approximation for different coupling regimes (weak or strong coupling) \cite{{key16}, {key19}, {key32}, {key33}, {key34}} and at different working points \cite{{key23}, {key35}}.

Moreover, self-assembled semiconductor quantum dots present the most suitable zero-dimensional structures for many applications in new devices and new areas such as quantum information processing \cite{{key36}, {key37}} and the optoelectronic devices \cite{key38}. A great deal of attention has been paid to these structures since many fundamental properties (quantized electronic states..) are size dependent in the nanometer range. Nevertheless, the fact that there is not enough studies dealing with the decoherence in semiconductors charge qubits due to RTN presents a challenge. The purpose of this paper is to extend these works and to investigate the telegraph noise effects on the entanglement through evaluating the populations, the entropy and the concurrence. Therefore, we apply our analysis, which is motivated by experiments proving the presence of RTN in semiconductor self-assembled quantum dots, to the dephasing of two semiconductor charge qubits \cite{{key15}, {key16}, {key17}}.

\vspace{1cm}

\qquad We consider a system with two interacting electrons in a symmetric coupled quantum dots at mutual distance 2d driving by an oscillatory electric field and coupled longitudinally to a classical telegraph noise,
\begin{equation}
   H=\sum_{i=1,2} h_i+ V_c + H_{RTN}
\end{equation}
 
here $h_i$ presents the single particle effective mass Hamiltonian in including the external electric field, $V_c=\frac{e^2}{k|r_1-r_2 |}$ is the Coulomb interaction and $ H_{RTN}$ corresponds to random telegraph noise $(RTN)$.
Let us start with a brief analyze of system without RTN. The two- dimensional single particle effective mass Hamiltonian of our system is given by:
\begin{equation}
 h_i (x,y)=\frac{P_i^2}{2m_i^*}+ex_i F(t)+V_{conf}
\end{equation}
here $ex$ is the dipole operator, the AC-electric field $F(t)$ is applied along the $x$ direction with the form: $ F(t)=F_0 cos⁡(wt)$. The electrons are confined by a Gaussian potential $V_{conf}(x,y)=-\frac{m^*\omega_0^2}{2}a_B^2(e^{-\frac{(x-d)^2+y^2 }{a_B^2}}+e^{-{\frac{(x+d)^2+y^2 }{a_B^2}}})$ in the quantum well plane.
\begin{figure}[h]
  \includegraphics[width=8cm]{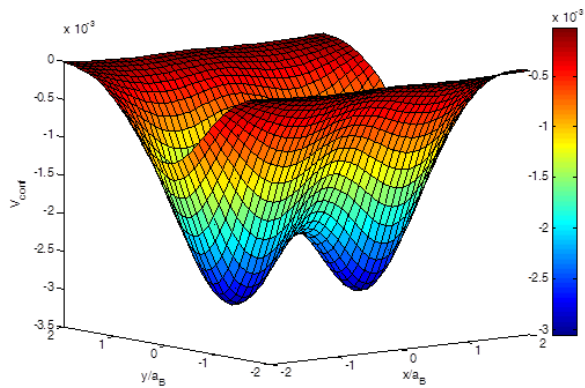}
  \caption{(Color online) The confinement potential scheme for fixed interdot distance $d=a_B$.}
  \label{fig1}
\end{figure}
We can use different material parameters values to reflect various self-assembled quantum dot systems. We take as material parameters for $GaAs$ quantum dot, $m^*=0.067m_e$ for the electron effective mass and $\epsilon_r=13.1$ for the dielectric constant. The electron mass is denoted $m_e$. The confinement strength is set to $\hbar\omega=6meV$ and $a_B=\sqrt{\frac{\hbar}{({m^*}\omega)}}$ being the effective Bohr radius.
The following investigation take as starting point the uncoupled dots at large distances, $d \gg a_B$. We have fixed the half interdot distance at $d=30nm$. By the choice of a large interdot distance, the decoherence due to the interaction with a bath of phonons can be neglected \cite{key8}.
The molecular orbital states of the field-free Hamiltonian are developed within the Hund-Mulliken approximation and with properly symmetrized products of Hermite functions that are the familiar solutions of the single-particle harmonic oscillator in two dimensions.
The symmetric and asymmetric basis functions correspond to singlet and triplet states, respectively. This basis has the advantage of yielding analytic expressions for the Hamiltonian matrix elements. A detailed investigation of the method can be found in a previous paper \cite{key39}. The spectrum and the eigenstates can be calculated by diagonalizing the two electrons hamiltonian in a truncated Hilbert space.
One natural approach to build a qubit is to use the different charges states, so that the charge qubit states can be defined trough the position of a single mobile electron. The resulting qubit is supposed to evolve in the basis spanned by the states $\left|0\right\rangle and \left|1\right\rangle$ which describe the electron localized in the right and the left dot respectively.
We consider an oscillatory electric field having an amplitude $F_0$ of about $0.4kV/cm$ and a frequency equal to that corresponding to the difference between the two lowest states $\hbar\omega_0=E_{s_2}-E_{s_1}$. It is interesting to note that the electric field allows the observation of coherent oscillations between the quantum states of the two qubits system which is necessary for their entanglement in the presence of a decoherence sources. Evidently, the latter does not mix singlet and triplet states, and thus the spin-triplet state is insensitive to the applied field within the truncated basis mentioned previously. Hence, we will focus our study on the three lowest singlets states $(E_{s_1},E_{s_2},E_{s_3})$. The latter states show a relatively simple and straightforward dependence on the confinement strength and the inter-dot distance. It appears that, at large inter-dot distance and for a system comprising two equal laterally-disposed dots, the energy spectrum presents two series of two nearly degenerate states. And hence, the dynamics is determined by transition between the ground state and the first excited state. In this case, we obtain a two level system. Furthermore, in the resonant case $\hbar\omega_0=E_{s_2}-E_{s_1}$, the population frequency matches the Rabi frequency $\Omega_r$ \cite{{key40}, {key41}}.

\vspace{1cm}

\qquad We turn now to the $RTN$ effects. It worth mentioning here that the classical telegraph noise is considered as a stochastic process produced by a two levels tunneling fluctuator. In this case, each process is represented by a function $\chi(t)$ switching randomly between two values, $\pm1$.
As we noted above, several mechanisms have been envisioned as $RTN$ sources, among them the dynamics of a background charges or a charge traps. Telegraph noise can be thought of as coming from a fluctuator with bi-stable states, $1$ and $2$, randomly fluctuating between them on time scales ranging from milliseconds to fractions of a nanoseconds.
 A fluctuator considered as a two level tunneling system is characterized by its coupling to the qubit $\nu$, and its switching rates $\gamma_{12}$ and $\gamma_{21}$ of the transition between its states $(1\rightarrow 2$ and $2\rightarrow 1)$. The coupling strength depends both on the working point of the qubit and the distance between the qubit and the fluctuator. These parameters vary independently in a wide energy range.
A charge, during its movement or a rearrangement of dynamic defects produces a randomly fluctuating (in time) electric field that can act upon the qubit and shift its charge states energies. Therefore, the environment effect is modeled classically through bringing into play a stochastic noise term in the system Hamiltonian as following:
\begin{equation}
H_{RTN}=\hbar\nu\chi(t)
\end{equation}
$\nu$ denotes the coupling strength between the fluctuator and the system in frequency units and the random telegraph noise sequence $\chi(t)$ switches instantly between $\pm1$, whose flipping events are Poisson distributed with an average switching rate $\gamma_{12}$ and $\gamma_{21}$. Without loss of generality, we can assume, for simplicity, a symmetric jump process with both rates of interstate $RTN$ states occurring with equal probability, i.e. $\gamma_{12}=\gamma_{21}=\gamma$. Indeed, since fluctuators with large level splitting are frozen in their ground states, only fluctuators with the energy splitting less than temperature contribute to the dephasing \cite{key18, key19, key25}.
At any given time $t$, the distribution of feasible values $RTN$ is evidently not a Gaussian. The probability for these jumps to occur in a given time-interval is assumed to be independent of the previous history of the process, i.e. the process is of the ‘Markov’ type.
The noise affects the qubit through a shift in the energy levels of the two states and hence introduces a random contribution to the relative phase of the two states. A qubit’s system undergoing these energy fluctuations will acquire a relative phase:
\newcommand{\deriv}{\mathrm{d}}
\begin{equation}
\varphi(t)=-\int_0^t \nu\chi (t^{'}) \, \mathrm{d}t^{'}
\end{equation}
The phase-memory functional describes the relative phase picked up during the time evolution by one state of a qubit relative to another. As expected, telegraph noise in our case leads to pure dephasing where the populations are preserved and only the off-diagonal elements of the density matrix (coherences) gain an additional coherence factor in their oscillatory time evolution.
\begin{equation}
\rho=\begin{pmatrix}
   \rho_{s_1s_1}(t) &\rho_{s_1s_2}(0) e^{-i\hbar\Omega_rt} D(t) \\
   \rho_{s_2s_1}(0) e^{+i\hbar\Omega_rt} D(t) &\rho_{s_2s_2}(t)
\end{pmatrix}
\end{equation}
The coherence factor is defined as the average of the phase-factor $e^{i\varphi(t)}$ .
\begin{equation}
D(t)=\left\langle e^{i\varphi(t)}\right\rangle
\end{equation}
Indeed, telegraph noise is an example of inherently non-Gaussian noise and the probability distribution of the phase has a pronounced non-Gaussian shape. Yet, it is worthwhile to investigate the Gaussian approximation, despite being not obvious in a general way, in order to determine the range of validity and applicability of the Gaussian assumption. Given that the time $t$ entering the phase exceeds the correlation time $\gamma^{-1}$ of the telegraph noise, the integral can be considered as the sum of a large number of uncorrelated contributions. Accordingly, by virtue of the central limit theorem, the phase $\varphi(t)$ is distributed according to a Gaussian. Indeed, the noise is completely defined by the correlation functions, one would get \cite{{key35}, {key36}} then:
\begin{equation}
D_{Gauss}(t)=exp⁡[-\frac{\nu^2}{2\gamma}(t-\frac{1}{2\gamma} (1-e^{-2\gamma t}))]
\end{equation}
However, the true telegraph noise induces non-Gaussian behavior; we assume already that such telegraph process does not feel any feedback. Therefore, the time evolution of the probability densities of finding this process in a state $+$ or $-$ with a value $\varphi$ is of ‘Markov’ type.
Taking the initial conditions into account, we obtain \cite{{key35}, {key36}}:
\begin{equation}
D_{Exact}(t)=\frac{1}{2} e^{-\gamma t} [(1+\frac{\gamma}{i\Omega}) e^{i\Omega t}+(1-\frac{\gamma}{i\Omega})e^{-i\Omega t}]
\end{equation}
With $\Omega=\sqrt{\gamma^2-\nu^2 }$.
\begin{figure}[ht]
  \includegraphics[width=9cm]{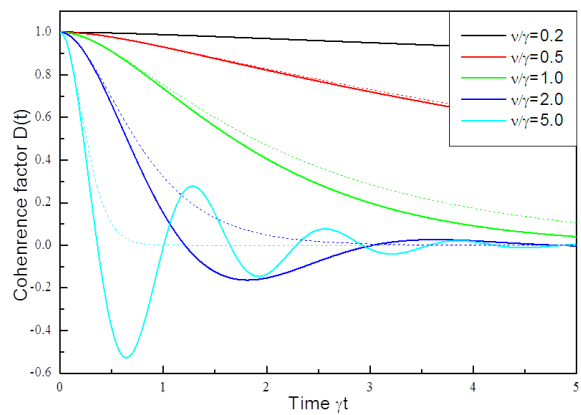}
  \caption{(Color online) Time evolution of the coherence factor $D(t)$ for different coupling values $\nu$ in the case of random telegraph noise. The dashed lines show the Gaussian approximation (see the text).}
  \label{fig2}
\end{figure}
Both of the two expressions show the dependence of the coherence factor upon the coupling strength $\nu$ and the switching rate $\gamma$.
The time evolution of the coherence factor corresponding to the Gaussian approximation and exact solution respectively are shown in Fig.\ref{fig2}. Both functions are plotted for different values of ratio $\nu / \gamma$. The decay of the coherence $D_{Gauss}(t)$ is monotonous and it doesn't have any zeros on the real axis. At long times $\gamma t\gg1$, the coherence factor decays exponentially with time $D_{Gauss}(t)\propto e^{(-\Gamma t)}$ at a decoherence rate $\Gamma=\frac{\nu^2}{4\gamma}$.
As it is displayed in the same figure, the exact coherence factor demonstrates qualitatively different features for small and large ratios $\nu / \gamma$ corresponding to weak and strong coupling respectively.
Fig.\ref{fig2} displays the appearance of coherence oscillations when $\nu>\gamma$ (strong coupling), as $\Omega$ becomes imaginary.
Comparing the two expressions we notice that the Gaussian result is a good approximation for the exact solution in the weak coupling limit $\nu\ll\gamma$. However, in the strong coupling case, when $\nu\ge\gamma$, the exact solution strongly differs from the Gaussian approximation. This latter, fails even qualitatively and drastically underestimates the phase memory functional. In the following, we will use the exact solution obtained above to analyze quantitatively the charge qubits decoherence.

\qquad We have evaluated the coherence factor related to a $RTN$ and we now turn to investigate $RTN$ effects on our two charge qubits. The more conventional methods to analyze the dynamics of the quantum system in the presence of decoherence source rely on studying the reduced density matrix.
We are interested in the time evolution of the two qubits system by density matrix $\rho(t)$, with an arbitrary initial condition at $t=0s$, where the corresponding system has been in its ground state corresponding to single dot occupancies. The time evolution of a general quantum system is governed by the Liouville-von Neumann equation:
\begin{equation}
i\hbar\rho(t)=[H,\rho(t)]
\end{equation}
\begin{figure}[ht]
  \includegraphics[width=9cm]{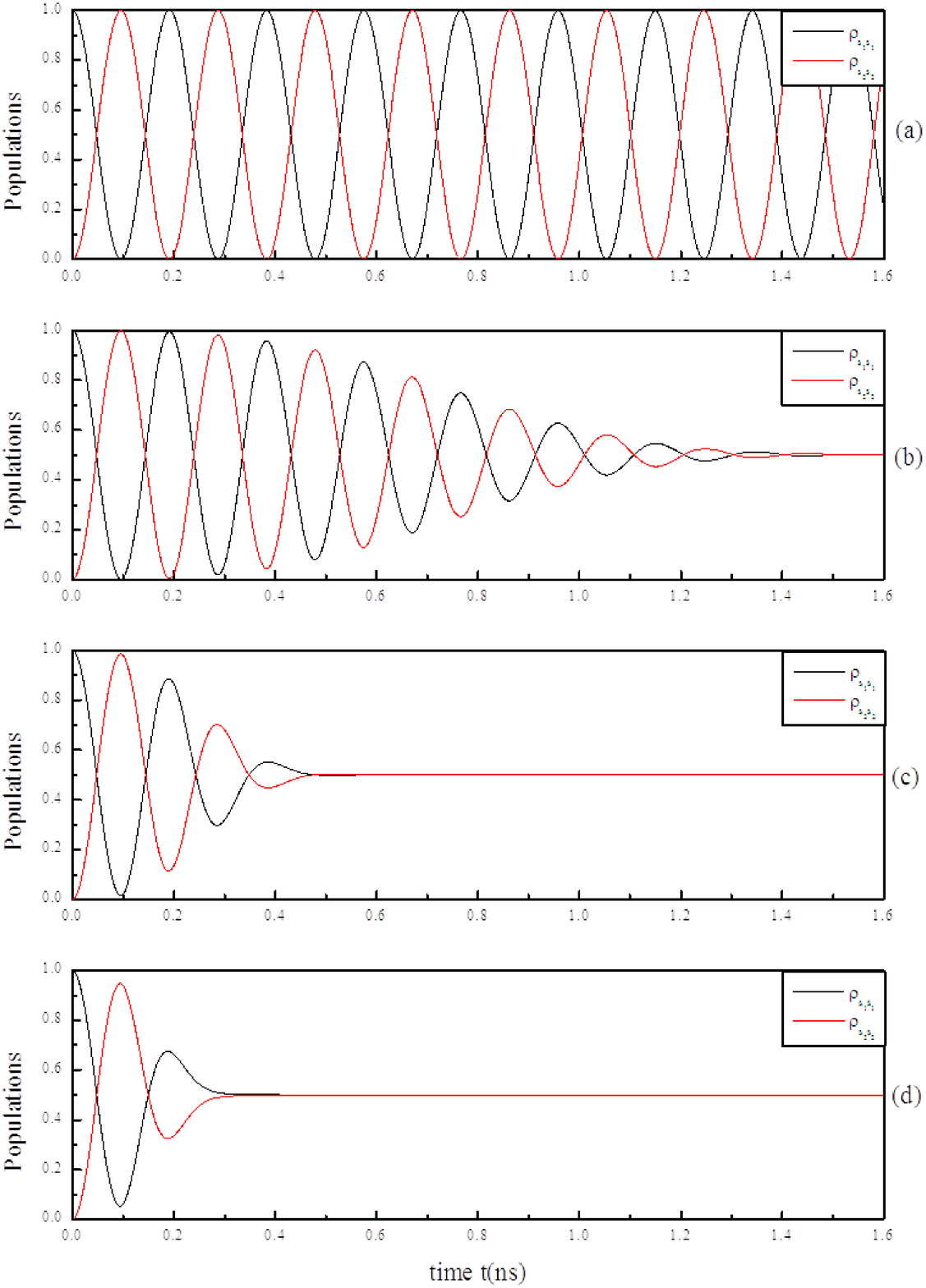}
  \caption{(Color online) Time evolution of the populations for different coupling values $\nu$ in the case of random telegraph noise $(RTN)$ with the ratio $\nu / \gamma=1.0$. (a) In the absence of $RTN$. (b) $\nu=10Mhz$. (c) $\nu=50Mhz$. (d) $\nu=100Mhz$.}
  \label{fig3}
\end{figure}
The Liouville-von Neumann equation allows us to exploit the intrinsic quantum decoherence effects in the electron charge coherency. For instance, as a result of coupling with the environment, the off-diagonal element of the density matrix decays as a function of time as described above.In order to investigate the effect of the RTN on the populations, arbitrary values of both $\nu$ and $\gamma$ have been used. Having examined the time dependencies of the populations shown in Fig.\ref{fig3}, we can conclude that the more important the coupling strength is, the more the random telegraph noise acts on the qubits dynamics and destructs the coherence between system states and vice versa.
It is clearly visible from Fig.\ref{fig3}.a that in absence of telegraph noise, the amplitude of the populations does not diminish at all and thus the qubits gets their initial entanglement back completely.
In the presence of random telegraph noise and by examining the Fig.\ref{fig3}, the populations show a damped oscillatory behavior. We attribute this effect to the competition between the electric field impact on the inter-qubit interactions and the telegraph noise decays. For longer time, the correlated decay becomes dominant and leads to a damped oscillatory decay of the entanglement.
The coherence phase can persist until $1.4ns$ for the low coupling strength corresponding to $\nu=10Mhz$. However, for a coupling strength $\nu=50Mhz$, the telegraph noise processes do not go over more than three Rabi cycles, with which a significant effect can be seen in maintaining the charge coherence, as it is exhibited below.
In others words, we show that even a low coupling strength among the system and telegraph noise may cause a substantial decrease of the populations amplitude. It is worth mentioning here that bearing in mind that the system entanglement lasts longer upon decreasing the coupling strength, we presume that the decoherence time can reach values higher than what we have found only with playing with the values of $\nu$ and $\gamma$.
As it is displayed in  Fig.\ref{fig3}, we can follow the evolution of each state population just within a brief time after which all the qubit states gain the same probability. In order to make the issue more clear we have evaluated the linear entropy as function of time \cite{key42}. The linear entropy provides a measure of the mixed character of the system described by a density matrix.
\begin{equation}
S(\rho)=tr(\rho-\rho^2)
\end{equation}
The entropy reaches its minimum value $0$ for a pure state system. Nonzero values of this quantity then provide a quantitative measure of the non-purity of the system states. We explore the influence of the coupling strength $\nu$ on $S$. In Fig.\ref{fig4} we plot the time evolution of the corresponding linear entropy for different values of coupling strength $\nu$.
\begin{figure}[ht]
  \includegraphics[width=9cm]{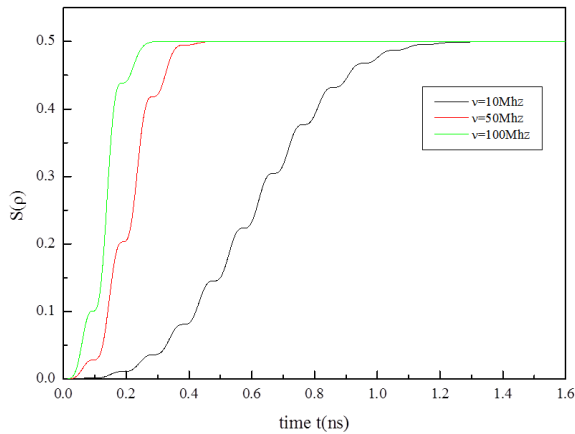}
  \caption{(Color online) Time evolution of the entropy for different coupling values $\nu$ in the case of random telegraph noise (RTN) with the ratio  $\nu / \gamma=1.0$.}
  \label{fig4}
\end{figure}
In the absence of $RTN$, we have a pure state with a linear combination of two qubits states $\left|00\right\rangle$, $\left|11\right\rangle$, $\left| 01\right\rangle$ and $\left|10\right\rangle$.
In Fig.\ref{fig4}, we show that starting from a pure state and in the presence of $RTN$ the state becomes mixed indicating that the noise quickly destroys the qubits states superpositions. We see that the linear entropy increases to a final stable value. In this case, all the possible two-qubit states have the same probability .i.e. the populations are equally distributed. The entropy of qubits increases due to the pure dephasing. The results reveal that the initial pure qubit states become mixed states. The stronger the coupling to the telegraph noise, the faster the purity system destruction.
Comparing the time evolution of the populations exhibited in Fig.\ref{fig3} for each coupling strength, the time at which the linear entropy reaches the final stable value is the same and matches the moment in which the populations becomes equal.

\begin{figure}[ht]
  \includegraphics[width=9cm]{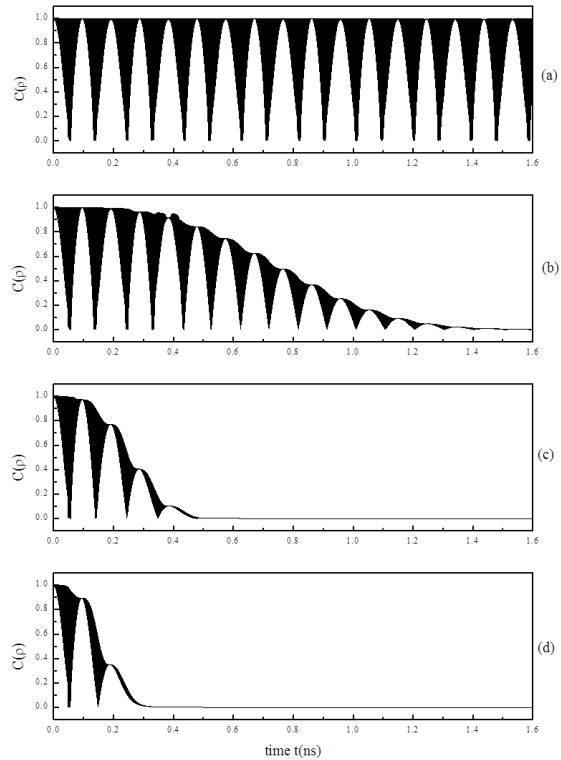}
  \caption{Time evolution of the concurrence for different coupling values $\nu$ in the case of random telegraph noise (RTN) with the ratio $\nu / \gamma=1.0$. (a) In the absence of $RTN$. (b) $\nu=10Mhz$. (c) $\nu=50Mhz$. (d)$\nu=100Mhz$.}
  \label{fig5}
\end{figure}

\qquad For a better understanding of the effect of interaction among the two qubits and the telegraph noise on decoherence, we must study the dynamics of two qubit entanglement. The entanglement for any bipartite system is often identified by examining and measuring the Wootters concurrence \cite{key43}. The concurrence varies from $0$ for the disentangled state to $1$ for the maximally entangled state. For any pair of qubits, the concurrence may be calculated explicitly from its density matrix $\rho$ for qubits $A$ and $B$. It is defined as
\begin{equation}
C^{AB}(t)=max\left|\sqrt{\lambda_1}-\sqrt{\lambda_2}- \sqrt{\lambda_3}-\sqrt{\lambda_4},0\right|
\end{equation}
with $\lambda_i$ are the square roots of the eigenvalues in decreasing order of the matrix $\rho_{AB}\tilde{\rho}_{AB}$ arranged.
\begin{equation}
\tilde{\rho}_{AB}=(\sigma_y\otimes\sigma_y)\rho_{AB}^* (\sigma_y\otimes\sigma_y)
\end{equation}
where $\rho_{AB}^*$ denotes the complex conjugation of $\rho_{AB}$ in the standard basis $\left|00\right\rangle$, $\left|11\right\rangle$, $\left| 01\right\rangle$ and $\left|10\right\rangle$ and $\sigma_y$ is the Pauli matrix expressed in the same basis.
Fig.\ref{fig5} shows the concurrence evolution as a function of time. One can clearly see the influence of the coupling strength on the concurrence and the qubits coherence phase. The most interesting observation is that, contrarily to the decoherence factor, the concurrence and the entanglement are independent of the ratio $\nu / \gamma$. Then, we can obtain the concurrence value according to the Gaussian approximation as the exact solution of $D(t)$ for any ratio $\nu / \gamma$.
In Fig.\ref{fig5}.a, we plot the dynamical evolution of entanglement when $\nu=0Mhz$, i.e in absence of $RTN$ and any environmental perturbation. This is an ideal case of close quantum systems whose dynamics is only influenced by the initial condition of the entangled qubits and the inter-qubit interactions due to electric field. The concurrence oscillates between $0$ and $1$ involving the evolution of the two charge qubits between maximally entangled and unentangled states. The inset of  Fig.\ref{fig5}.a shows the long time behavior of entanglement for the latter case. We note that for selected times, we have a maximally entangled state .i.e. $C=1$ corresponding to the single dot occupancy $(\rho_{s_1s_1}=1)$ or the double dot occupancy $(\rho_{s_2s_2}=1)$, and a disentangled state $C=0$ corresponding to $\rho_{s_1s_1}=\rho_{s_2s_2}=0.5$.
As expected for the other case, telegraph noise affects the oscillations and destructs the entanglement of the system. At a given finite time, these oscillations vanish completely. We see that the entangled qubits gets repeatedly disentangled and entangled leading to periods in the concurrence. The oscillation amplitude decreases  with time down to a complete vanishing after a relatively long time. We have found that this trend in entanglement also holds for other values of the parameter $\nu$ as well. As expected, the destruction of the coherence becomes increasingly important upon increasing the coupling strength $\nu$.
According to Fig.\ref{fig3} and  Fig.\ref{fig5}, we note that the time when the two qubit states gain the same probability  $\rho_{s_1s_1}=\rho_{s_2s_2}=0.5$ is almost equal to that corresponding to $C=0$. From this time on, the probability of finding the two electrons in the same dot will be equal to that relative to the finding of one electron in each dot.

\section*{Conclusions}
\qquad In this paper, for a set of initial two qubit states driven by an oscillatory electric field, we have investigated quantum entanglement decay due to interaction with classical telegraph noise. We have shown that noisy classical environments may completely destroy the coherences between two qubits. To this end, we have considered the time evolution of the degree of entanglement, as measured by the populations, the linear entropy and Wootters concurrence.  We found that $RTN$ completely destroys the coherences between two qubits. This effect is more important with increasing the coupling strength between qubits and $RTN$.

\section*{ACKNOWLEDGMENTS}
We are grateful to R. Ferreira for useful discussions.


\vspace{1cm}

\begin{thebibliography}{99}

\bibitem{key1}J. R. Petta, A. C. Johson, J. M. Taylor, E. A. Laird, A. Yacoby, M. D. Lukin, C. M. Marcus, M. P. Hanson and A. Gossard, Science \textbf{309}, 2180 (2005).

\bibitem{key2}D. J. Mowbray, and M. S. Skolnick, J. Phys. D: Appl. Phys \textbf{38}, 2059–2076 (2005).

\bibitem{key3}K. D. Petersson, J. R. Petta, H. Lu, and A. C. Gossard, Phys. Rev. Lett \textbf{105}, 246804 (2010).

\bibitem{key4}D. Loss and D. P. DiVincenzo, Phys. Rev. A \textbf{57}, 120 (1998).

\bibitem{key5}T. Hayashi, T. Fujisawa, H. D. Cheong, Y. H. Jeong, and Y. Hirayama, Phys. Rev. Lett \textbf{91}, 226804 (2003).

\bibitem{key6}R. Nepstad, L. Sælen, I. Degani, and J. P. Hansen. J. Phys.: Condens. Matter \textbf{21}, 215501 (2009).

\bibitem{key7}L. Fedichkin, and A. Fedorov, Phys. Rev. A \textbf{69}, 032311 (2004).

\bibitem{key8}W. Ben Chouikha, S. Jaziri and R. Bennaceur, Phys. Rev. A \textbf{76}, 062303 (2007).

\bibitem{key9}S. D. Barrett, and G. J. Milburn, Phys. Rev. B \textbf{68}, 155307 (2003).

\bibitem{key10}A. Shnirman, G.Schon, and Z. Hermon, Phys. Rev. Lett \textbf{79}, 2371 (1997).

\bibitem{key11}Y. A. Pashkin, O. Astafiev, T. Yamamoto, Y. Nakamura, and J. S. Tsai, Quantum Inf. Process \textbf{8}, 55 (2009).

\bibitem{key12}M.-E. Pistol, Phys. Rev. B \textbf{63}, 113306 (2001).

\bibitem{key13}N. Panev, M.-E. Pistol, V. Zwiller, L. Samuelson, W. Jiang, B. Xu, and Z. Wang, Phys. Rev. B \textbf{64}, 045317 (2001).

\bibitem{key14}N. Panev, M.E. Pistol, and S. Jeppesen. V. P. Evtikhiev, A. A. Katznelson, and E. Yu. Kotelnikov. Journal of Applied Physics \textbf{92}, 12 (2002).

\bibitem{key15}J. Bergli, Y. M. Galperin, and B. L. Altshuler, Phys. Rev. B \textbf{74}, 024509 (2006).

\bibitem{key16}Y. M. Galperin, B. L. Altshuler, J. Bergli, and D. V. Shantsev. Phys. Rev. Lett \textbf{96}, 097009 (2006).

\bibitem{key17}O. P. Saira, V. Bergholm, T. Ojanen, and M. Mottonen, Phys. Rev. A \textbf{75}, 012308 (2007).

\bibitem{key18}R.W. Simmonds, K.M. Lang, D. A. Hite, S. Nam, D. P. Pappas, and J. M. Martinis, Phys. Rev. Lett \textbf{93}, 077003 (2004).

\bibitem{key19}E. Paladino, L. Faoro, G. Falci, and Rosario Fazio. Phys. Rev. Lett \textbf{88}, 228304 (2002).

\bibitem{key20}I. V. Yurkevich, J. Baldwin, I. V. Lerner, and B. L. Altshuler, Phys. Rev. B \textbf{81}, 121305(R) (2010).

\bibitem{key21}T. Itakura, and Y. Tokura, Phys. Rev. B \textbf{67}, 195320 (2003).

\bibitem{key22}Y. M. Galperin, B. L. Altshuler, J. Bergli, D. Shantsev, and V. Vinokur, Phys. Rev. B \textbf{76}, 064531 (2007).

\bibitem{key23}E. Paladino, A. Mastellone, A. D’Arrigo, and G. Falci. Proceedings of the international Symposium on Mesoscopic Superconductivity and Spintronics. Atsugi, Japan (2004).

\bibitem{key24}\L{}. Cywi\'{n}ski, R. M. Lutchyn, C. P. Nave, and S. DasSarma, Phys. Rev. B \textbf{77}, 174509 (2008).

\bibitem{key25}G. Falci, A. D’Arrigo, A. Mastellone, and E. Paladino. Phys. Rev. Lett \textbf{94}, 167002 (2005).

\bibitem{key26}G. Burkard, Phys. Rev. B \textbf{79}, 125317 (2009).

\bibitem{key27}G. Ithier, E. Collin, P. Joyez, P. J. Meeson, D. Vion, D. Esteve, F. Chiarello, A.Shnirman, Y. Makhlin, J. Schriefl, and G. Sch\"{o}n, Phys. Rev. B \textbf{72}, 134519 (2005).

\bibitem{key28}A. J. Leggett, Rev. Mod. Phys \textbf{59}, 1 (1987).

\bibitem{key29}U. Weiss, Quantum Dissipative Systems (Word Scientific, Singapore, 1999), 2nd ed.

\bibitem{key30}F. Yoshihara, K. Harrabi, A. O. Niskanen, Y. Nakamura, and J. S. Tsai, Phys. Rev. Lett \textbf{97}, 167001 (2006).

\bibitem{key31}H. Gutmann, F. K. Wilhelm, W. M. Kaminsky, and S. Lloyd, Phys. Rev. A \textbf{71}, 020302(R) (2005).

\bibitem{key32}C. Neuenhahn, B. Kubala, B. Abel, and F. Marquardt, Phys. Status Solidi B \textbf{246} 5,1018-1023 (2009).

\bibitem{key33}B. Abel and F. Marquardt, Phys. Rev. B \textbf{78}, 201302 (R) (2008).

\bibitem{key34}A. Berthelot, C. Voisin, C. Delalande, Ph. Roussignol, R. Ferreira, and G. Gassabois, Advances in Mathematical Physics \textbf{10}. 1155. 494738 (2010).

\bibitem{key35}J. Bergli, Y. M. Galperin, and B. L. Altshuler, New Journal of Physics \textbf{11}, 025002 (2009).

\bibitem{key36}A. Imamoglu, D. D. Awschalom, G. Burkard, D. P. DiVincenzo, D. Loss, M. Sherwin, and A. Small. Phys. Rev. Lett \textbf{83}, 4204–4207 (1999)

\bibitem{key37}A. Zrenner, E. Beham, S. Stufler, F. Findeis, M. Bichler, and G. Abstreiter. Nature \textbf{418}, 612 (2002).

\bibitem{key38}A. R. Kovsh, N. A. Maleev, A. E. Zhukov, S. S. Mikhrin, A. P. Vasil’ev, E. A. Semenova, Yu. M. Shernyakov, M. V. Maximov, D. A. Livshits, V. M. Ustinov, N. N. Ledentsov, D. Bimberg, and Zh. I. Alferov. Journal of Crystal Growth \textbf{251}, 729–736 (2003).

\bibitem{key39}G. Burkard, D. Loss, and D. P. DiVincenzo, Phys. Rev. B \textbf{59}, 2070 (1999).

\bibitem{key40}H. Htoon, T. Takagahara, D. Kulik, O. Baklenov, A. L. Holmes, Jr., and C. K. Shih. Phys. Rev. Lett \textbf{88}, 087401 (2002).

\bibitem{key41}P. Borri, W. Langbein, S. Schneider, and U. Woggon. Phys. Rev. B \textbf{66}, 081306 (2002).

\bibitem{key42}W. H. Zurek, S. Habib, and J. P. Paz, Phys. Rev. Lett \textbf{70}, 1187 (1993).

\bibitem{key43}S. Hill and W. K. Wootters, Phys. Rev. Lett \textbf{78}, 5022 (1997).

\end{thebibliography}
\end{document}